\RequirePackage{fix-cm}
\documentclass[twocolumn,epjc3]{svjour3}  
\smartqed  
\RequirePackage{graphicx}
\usepackage{amssymb}
\usepackage{amsmath}
\usepackage{epstopdf}
\RequirePackage{float}
\usepackage{bbm}
\usepackage{mathrsfs}
\usepackage{xcolor}

\newcommand{\eq}[2]{\begin{align}\label{#1}#2\end{align}}

\newcommand{\ben}{\begin{enumerate}}\newcommand{\een}{\end{enumerate}}
\newcommand{\Ref}[1]{(\ref{#1})}

\newcommand{\td}{thermodynamic~}

\newcommand{\al}{\alpha}

\renewcommand{\td}{thermodynamic~}

		\newcommand{\der}[3]{\frac{d^#3 #1}{d#2^#3}}
		
		\newcommand{\pder}[3]{\frac{\partial ^#3 #1}{\partial #2^#3}}
		\newcommand{\pderuno}[2]{\frac{\partial  #1}{\partial #2}}
		
		\newcommand{\bea}{\begin{eqnarray}} 
		\newcommand{\eea}{\end{eqnarray}}
\journalname{Eur. Phys. J. C}
\begin{document}

\title{Free energy and entropy for finite temperature quantum field theory under the influence of periodic backgrounds
}


\author{M. Bordag\thanksref{e1,addr1}
        \and
        J. M. Mu\~noz-Casta\~neda\thanksref{e2,addr2} 
        \and
        L. Santamar\'\i  a-Sanz\thanksref{e3,addr2}
}

\thankstext{e1}{e-mail: bordag@uni-leipzig.de}
\thankstext{e2}{e-mail: jose.munoz.castaneda@uva.es}
\thankstext{e3}{e-mail: lucia.santamaria@uva.es}


\institute{Institut f\"ur Theoretische Physik, Universit{\"a}t Leipzig, 04103 Leipzig (GERMANY) \label{addr1}
           \and
           Departamento de F\'{\i}sica Te\'orica, At\'omica, y \'Optica, Universidad de Valladolid, 47011 Valladolid (SPAIN) \label{addr2}
}

\date{Received: date / Accepted: date}

\maketitle

\begin{abstract}
The basic thermodynamic quantities for a non-interacting scalar field in a periodic potential composed of either a one-dimensional chain of Dirac $\delta$-$\delta^\prime$ functions or a specific potential with extended {\color{black} compact} support
are calculated.   First, we consider the representation in terms of real frequencies (or one-particle energies). Then we turn the axis of frequency integration towards the imaginary axis by a finite angle, which allows for easy numerical evaluation, and finally turn completely to the  imaginary frequencies and derive the corresponding Matsubara representation, which this way appears also for systems with band structure. In the limit case $T \to 0$ we confirm earlier results on the vacuum energy. We calculate for the mentioned examples the free energy and the entropy and generalize earlier results on negative entropy.
{\color{black}}
\end{abstract}

\section{Introduction}
Since the seminal work by H. B. G. Casimir \cite{cas48} and the experimental confirmation by Sparnay \cite{spa57,spa58} the theory of quantum fields interaction with classical backgrounds mimicking macroscopical objects has been a very active field of research (see Refs. \cite{Grib1994,milt-book,bord-book} and references therein). Most of the results obtained have focused on the study of the dependence of the zero-temperature quantum vacuum energy and its sign with the geometry (see Refs. \cite{emig-prl07,emig-prd08,rahi-ped09,kenneth-prb08,kenneth-prl06}).  In the last decade the use of boundary conditions allowed by the principles of quantum field theory has been used to study the properties and sign of the quantum vacuum energy. {\color{black}In particular general boundary conditions were used to mimic idealised models of two plane parallel plates with arbitrary physical properties and topology changes (see Refs. \cite{asorey-npb13,asorey-jpa06,muno15-91-025028,munoz-lmp15}). }

Nearly 15 years ago in Casimir effect investigations the occurrence of negative entropy was noticed in \cite{geye05-72-022111}. In fact, \td puzzles were observed earlier, see, e.g., \cite{thir70-235-339}. However, not much attention was paid to \cite{geye05-72-022111} since there only the separation dependent part of the entropy was considered and the focus was on another, possibly related, effect; namely a violation of the Nernst's heat theorem while using the Drude model for the dielectric slab. This problem remains still unresolved and it is related to the choice of plasma or Drude model in Casimir force calculations, see \cite{lium19-100-081406} for the actual status.

{\color{black}In the last three years}, the entropy of Casimir effect related configurations was calculated for quite a large number of model systems. For three dimensional ones, a plasma plane or a plasma sphere, in \cite{milt17-96-085007,li16-94-085010} and \cite{bord18-51-455001,bord18-98-085010} and for some simple one dimensional examples in \cite{bord1807.10354}. In all these cases the single standing objects were considered and the complete entropy, except the black body part (contributions from the empty space), was computed. Again, negative entropy was observed. Now, one could speculate that negative entropy, like negative specific heat, could signal some instability of the system as discussed, for example, in \cite{thir70-235-339}. However, that is beyond the scope of the present paper.

The present paper is a continuation of the above line of research concerning entropy in simple systems, now on periodic background fields. We will develop a general formalism to compute finite temperature corrections to the quantum vacuum energy as well as the entropy for periodic classical background potentials represented by infinite chains of potentials with compact support. As an application of the formalism developed we will firstly study the free energy and entropy in a one dimensional lattice of delta functions, generalized to include derivative of delta function,
\eq{1}{V(x)=\sum\left(w_0\delta(x-an)+2w_1\delta'(x-an)\right).
}
In fact, {\color{black}the potential $w_0\delta(x)+2w_1\delta'(x)$} is a self-adjoint {\color{black} extension of the free particle hamiltonian on $\mathbb{R}/\{0\}$ that generalises} the simple delta function and in place of \Ref{1} we will use the corresponding matching conditions defined in Refs. \cite{muno15-91-025028,gadella-pla09}. This model can also be viewed as a version of the much studied Kronig-Penney model and its pleasant feature is the possibility to work with mostly explicit formulas, showing nevertheless the interesting features we are interested in. At zero temperature, in \cite{bord19-7-38} the vacuum energy was calculated for this model. Some formulas from that paper will prove to be useful below. Secondly we will apply our general formulas to the case of a periodic potential built from as an infinite array of P\"oschl-Teller potentials modulated by Heaviside functions as in Ref. \cite{guil11-50-2227}.

The paper is organized as follows. In the next section we collect the necessary formulas for a generic periodic potential and the basic \td formulas. In section 3 we derive general representations of the free energy and entropy for arbitrary temperature which are convenient for the numerical evaluation. In section 4 we use the general formulas from section 3 to compute numerically the free energy and entropy for the two particular cases mentioned above. Finally in section 5 we present or concluding remarks.
\\Throughout the paper we will use a system of units where $\hbar=c=k_{\rm B}=1$.

\section{The model}
We consider the action of a massless scalar field $\phi(x)$ in (1+1)-dimensions,
\begin{equation}
S(\phi)= \frac{1}{2} \int dx^{1+1} \left[ (\partial \phi)^2 - U(x) \phi^2 \right],
\end{equation}
where the considered general background periodic potential reads as
\begin{eqnarray}
&&U(x)= \sum_{n\in \mathbb{Z}} V(x-na), \qquad a>0,\nonumber\\
&& V(x) \left\{
	       \begin{array}{ll}
		 \neq 0      & \mathrm{if\ } |x|\leq \epsilon{\color{black} /2},  \\
		 = 0 & \mathrm{if\ } |x|> \epsilon{\color{black} /2},
	       \end{array}
	     \right. \qquad \, \, \epsilon \leq a.\label{eqq3}
\end{eqnarray}
This scalar field $\phi(x)$ obeys the Schr\"odinger equation, after Fourier transform
\begin{equation}
\left( -\pder{}{x}{2} + {\color{black} U}(x)\right) \phi(x)= \omega^2 \phi(x)
\end{equation}
where $\omega=k$ are the frequencies of the quantum field modes.
Given the one particle states hamiltonian,
\begin{equation}\label{eq4}
\hat{K} = -\der{}{x}{2} + U(x),
\end{equation}
the band spectrum of the lattice can be written in terms of {\color{black}the transmission amplitude $t(k)$, and the reflection amplitudes $r_L(k)$ and $r_R(k)$ for the Hamiltonian
\begin{equation}
\hat{H}_V = -\der{}{x}{2} + V(x).
\end{equation}
It is of note that since $V(x)$ has compact support all the transmission amplitudes admit analytical continuation to the whole complex $k$-plane with a finite number of poles, i. e. $t(k)$, $r_L(k)$ and $r_R(k)$ are meromorphic functions over the complex $k$-plane (see Ref. \cite{galin-b}) }.
  The bands are determined by the real solutions of the spectral equation, i.e. by the zeroes of the {\color{black}meromorphic function}
\eq{eq6}{
&& f_{\theta}(k) = \cos(\theta) - h_V(k), \qquad \theta \in [-\pi/\pi],
}
where we {\color{black}define $h_V(k)$ as} \cite{gad1909.08603}
\eq{6a}{h_V(k)= \frac{1}{2t(k)} \left[ e^{-ika} + e^{ika} (t^2(k)-r_R(k)r_L(k))\right].
}
The parameter $\theta$ is {\color{black}connected with the quasi-momentum $q$ following from the Bloch periodicity,
\begin{equation}
\phi(x+a)=e^{iqa} \phi(x),
\end{equation}
by means of  $\theta=-qa$. }
For the definition of the thermodynamic quantities we put the model into a large box, $x\in [-L/2, L/2]$ with $ L \to \infty$ in the thermodynamic limit. Now the energies are discrete and the equation $f_\theta(k)=0$ turns into
\begin{equation}
\cos\left(\frac{i \pi }{N}\right)- h_V(\omega_{n,i})=0
\end{equation}
Here $N=L/a$ is the number of $V(x)$  potentials given by \eqref{eqq3} {\color{black} contained in the box $[-L/2, L/2]$}, $i$ labels the energy levels inside a group of levels which turns into a band for $L \to \infty$ and $n$ numbers the band. One of the  thermodynamic quantities that we need is the free energy,
\begin{equation}
\mathcal{F}=E_0+ \bigtriangleup_T \mathcal{F},
\end{equation}
where
\begin{equation}
E_0= \frac{1}{2} \sum_{n,i} \omega_{n,i}^{1-2s}
\end{equation}
is the vacuum energy (we have introduced the zeta regularization \cite{klaus-book}) and
\begin{equation}
\bigtriangleup_T \mathcal{F}= T \sum_{n,i} \, \textrm{ln} (1-  \, e^{-{\omega_{n,i}}/T}),
\end{equation}
is the temperature dependent part of the free energy.   The entropy $S$ follows with
\begin{eqnarray}
S&=&-\pderuno{F}{T},
\end{eqnarray}
which is the well know \td definition.

\section{Basic free energy and entropy formulas}
The general form of the band equation in terms of scattering coefficients
($t, r_R, r_L$) and the quasi-momentum $q$ in the first Brillouin zone  for the compact supported potential from which the comb is built is given by (\ref{eq6}). Since the cosine of the left hand side of (\ref{eq6}) is a bounded function, the energy spectrum of the system is organized into allowed/forbidden energy bands/gaps. The crystal spectrum will be obtained as
\begin{equation}
\textrm{spec}(\hat{K})= \lbrace \omega / f_\theta(\omega)=0\rbrace_{\theta \in [-\pi,\pi]}
\end{equation}
being $\hat{K}$ the quantum Hamiltonian that characterises the one particle states of the theory (\ref{eq4}) and $f_\theta(\omega)$ the spectral function given by (\ref{eq6}). The spectrum of the comb, $\textrm{spec}(\hat{K})$, is a band spectrum that depends on a continuous parameter $\theta$ and we will assume that there are no negative energy bands. Furthermore, for each $\theta$ fixed within the interval $[-\pi,\pi]$, $\textrm{spec}(\hat{K})_\theta$ is a discrete point spectrum. Hence, we can obtain the whole spectrum $\textrm{spec}(\hat{K})$ as the union of the 1-parameter point spectra $\textrm{spec}(\hat{K})_\theta$ for $\theta\in [-\pi,\pi]$  (see Ref. \cite{bord19-7-38} for details).

The temperature dependent part of the free energy can be computed as the sum of the Boltzmann factors
\eq{16a}{
B(\omega,T)&=T\ln \left[ 1- \textrm{exp}\left(-{\omega}/T\right) \right],
}
over the quantum field modes that form the comb spectrum,
\eq{16}{
\bigtriangleup_T \mathcal{F}&= \sum_{\omega\,  \in\,  \textrm{spec}(\hat{K})} B(\omega, T).
}
Here $\omega=k$ is the energy of the one-particle states of the quantum field theory. 

\begin{figure}[h]
\centering
\includegraphics[width=0.45\textwidth]{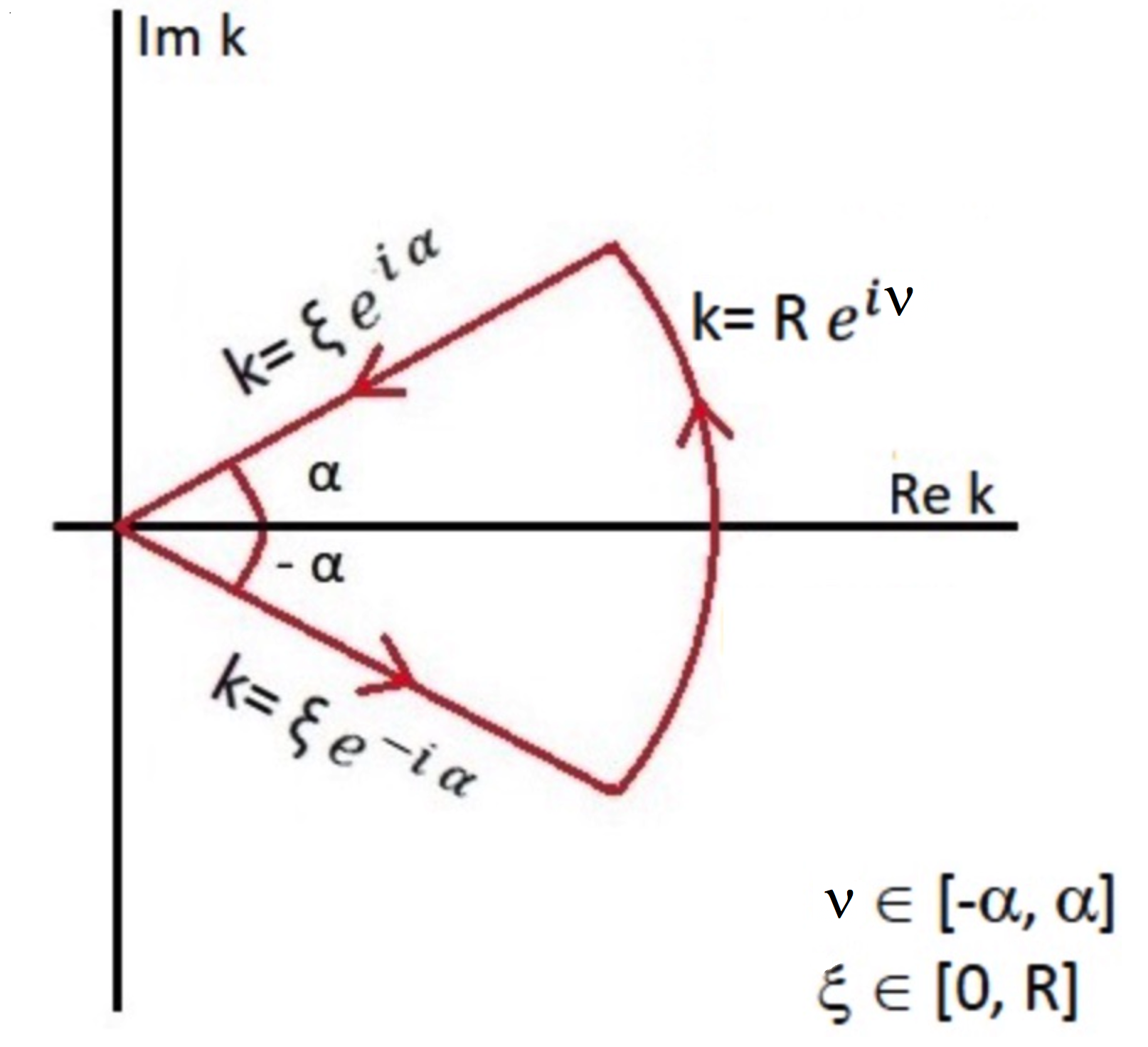}
\caption{\small Complex contour that encloses all the zeroes of $f_\theta(k)$. We take $R \to \infty$ and $\alpha$ a constant angle.}
\label{fig:contour}
\end{figure}

Now, the summation over the whole spectrum is equivalent to the summation over the spectrum for a fixed $\theta$ and then integrating the continuous parameter $\theta$ in $[-\pi, \pi]$. Adding over the spectrum is the summation over the zeroes of the secular function $f_\theta(\omega)$, i.e, we have to sum up the energies of each band for all the bands that form the whole spectrum,
\begin{equation}\label{eq16}
\bigtriangleup_T \mathcal{F}= \!\!\!\!\sum_{\omega\,  \in\,  \textrm{spec}(\hat{K})} \!\!\!\!\!B(\omega, T) =\!\! \int _0^\pi \frac{d\theta}{\pi}\!\!\! \sum_{\omega\,\in\, \textrm{spec}(\hat{K})_\theta}\!\!\!\!\! B(\omega, T).
\end{equation}
Since $\theta$ represents the quasi-momentum in the first Brillouin zone, integrating over the energies of each band is equivalent to integrating the quasi-momentum in the primitive cell. The summation over the zeroes of $f_\theta(k)$ (which give us the band structure) can be written down by using a complex contour integral (through the Cauchy integral formula) which involves the logarithmic derivative of the secular equation,
\begin{eqnarray}\label{eq14a}
\bigtriangleup_T \mathcal{F}= \int_{0}^{\pi} \frac{d\theta}{\pi} \oint_\Gamma \frac{dk}{2\pi i} B(k,T) \partial_k \log f_\theta(k),
\end{eqnarray}
where the contour  $\Gamma$  is represented in Figure \ref{fig:contour}.
The Boltzmann factors, $B(k,T)$, have a discrete set of branch points on the imaginary axis. The  integral ($\ref{eq14a}$) is well defined because $f_\theta(k)$ is an holomorphic function on $k$ and the logarithmic derivative of the secular equation has poles at the zeroes of $f_\theta(k)$ (which are the bands in the real axis when we sum over the quasi-momentum) and the residue coincides with the multiplicity of the corresponding zero.
When the variable $R$ tends to infinity, the integral over the circumference arc of the contour Figure \ref{fig:contour}  goes to zero {\color{black} since\footnote{{\color{black}From Ref. \cite{galin-b} the asymptotic behavior of the scattering amplitudes in this case is $$t(\vert k\vert\to\infty)\to1,\quad r_{R,L}(\vert k\vert\to\infty)\to0.$$ Therefore $h_V(\vert k\vert\to\infty)\to\cos(k a)$.}} 
\begin{equation*}
\lim_{R\to\infty}\cot(aR e^{\pm i \nu})B(R e^{\pm i\nu},T)=0,
\end{equation*}
for any $\nu\in(0,\pi/2)$.} Hence, integrating over the whole contour is equivalent to integrating over the two straight lines  $k=\xi e^{i\alpha}$ and $k=\xi e^{-i\alpha}$ being $\alpha$ a constant angle. In such a way $\bigtriangleup_T \mathcal{F}$ reads as
\begin{eqnarray}\label{eq18}
&&\bigtriangleup_T \mathcal{F}=\int_{0}^{\pi} \frac{d\theta}{\pi} \left[ - \int_0^\infty \frac{d\xi}{2\pi i} B(\xi e^{i\alpha},T) \partial_\xi \log f_\theta(\xi e^{i\alpha}) +\right.\nonumber\\
&&\left. \int_0^\infty \frac{d\xi}{2\pi i} B(\xi e^{-i\alpha},T) \partial_\xi \log f_\theta(\xi e^{-i\alpha})\right]
\end{eqnarray}
The residue theorem ensures that the result of this integration does not depend on the angle $\alpha$ taken in the contour. Furthermore, the complex contour chosen avoids the possible poles in the real axis and the pure imaginary axis of the functions that form the integrand. From formula (\ref{eq18}) we obtain
\begin{eqnarray}\label{eq19}
&&\bigtriangleup_T \mathcal{F}=\int_{0}^{\pi} \frac{d\theta}{\pi} \int_0^\infty \frac{d\xi}{2\pi i}  \left[ - B(\xi e^{i\alpha}, T) \partial_\xi \log f_\theta(\xi e^{i\alpha}) +\right.\nonumber\\
&&\left. B(\xi e^{-i\alpha}, T) \partial_\xi \log f_\theta(\xi e^{-i\alpha})\right]
\end{eqnarray}
But taking into account (\ref{eq6}) we can exchange the integrals in order to do $\theta$ integration first
\begin{eqnarray}
&&\int_0^\pi \frac{d\theta}{\pi} \partial_\xi \, \log f_\theta(\xi \, e^{\pm i\alpha}) = \int_0^\pi \frac{d\theta}{\pi} \frac{-\partial_\xi\,  h_V(\xi \, e^{\pm i\alpha})}{\cos \theta - h_V(\xi \, e^{\pm i\alpha})}=\nonumber\\
&&= \frac{ -\partial_\xi\,  h_V(\xi \, e^{\pm i\alpha})}{\sqrt{-1+h_V^2(\xi\,  e^{\pm i\alpha})}}= D_V(\xi\,  e^{\pm i\alpha})
\end{eqnarray}
Plugging this result in (\ref{eq19}) we obtain the final expression
\eq{eq21a}{
\bigtriangleup_T \mathcal{F}&=\int_0^\infty \frac{d\xi}{2\pi i}  \left[ - B(\xi e^{i\alpha}, T) D_V(\xi\,  e^{ i\alpha}) +\right.\nonumber\\
&~~~~~~~~~~~~~~~~\left. B(\xi e^{-i\alpha}, T) D_V(\xi\,  e^{- i\alpha})\right].
}
This formula can be applied to any comb whose individual potential of the primitive cell has a compact support not exceeding the lattice spacing.
It has the  advantage that it avoids possible oscillations of the integrand caused by the secular function on the real axis. Also it avoids the branch points on the imaginary axis. At once, for finite slope $\alpha$, the integrand has an exponential decrease which makes numerical evaluation easier.

\subsection{Free energy: real frequencies}
Another approach to compute the temperature dependent part of the free energy  is to work on the real line. In order to compute $\bigtriangleup_T \mathcal{F}$ we start from (\ref{eq16}) where we have to sum up the Boltzmann factors over the spectrum, or equivalent, to sum over the zeroes of the secular equation $f_\theta(\omega)=0$, which gives us the band energy structure of the comb:
\begin{equation}\label{eq22}
f_\theta(\omega)= \cos \theta - h_V(\omega)=0, \qquad \theta \in [-\pi/\pi].
\end{equation}
The allowed $\omega$ of the spectrum are given by the condition 
\begin{equation}\label{eq23}
|h_V(\omega)|\leq 1.
\end{equation}
Plugging (\ref{eq22}) into the following form
\begin{equation}
\theta(\omega)= \arccos h_V(\omega),
\end{equation}
the condition (\ref{eq23}) reads as
\begin{itemize}
\item Allowed $\omega$ $\quad \rightarrow \quad$ $\theta(\omega)=\arccos h_V(\omega) \in \mathbb{R},$
\item Forbidden $\omega$ $\, \rightarrow \quad$ $\theta(\omega)= \arccos h_V(\omega) \in \mathbb{I}$.
\end{itemize}
Hence, if we want to integrate the energy $\omega$ from the minimum energy to the maximum energy of each band, we have to make the change of variables $\theta \to \omega(\theta)$ and introduce the Jacobian of this transformation
\begin{eqnarray}\label{eq25}
\bigtriangleup_T \mathcal{F}&=&  \sum_n \int_{\omega_n(0)}^{\omega_n(\pi)} \frac{d\omega}{ \pi}  \pderuno{\theta}{\omega} B(\omega,T)
\end{eqnarray}
where $n\in \mathbb{N}$ indexes the bands. From (\ref{eq22}) we get
\begin{equation}
\pderuno{\omega}{\theta}= \frac{-\sin \theta}{\partial_\omega h_V(\omega)}.
\end{equation}
This result implies that, for each band that forms the spectrum, $\omega_n(\theta)$ is a monotone function between $\theta=0$ and $\theta=\pi$ {\color{black}(see Ref. \cite{gad1909.08603} for a detailed demonstration)}. Furthermore, for those extreme values of the quasi momentum $\theta=0,\pi$ in the first Brillouin zone there is always a maximum or a minimum of the band. We can distinguish two cases
\begin{itemize}
\item If $\omega_n(0)=\omega_{min}$  and $\omega_n(\pi)=\omega_{max}$  of the band
\begin{equation}
\pderuno{\theta}{\omega}=\left| \pderuno{\theta}{\omega}\right| >0,\qquad \forall \omega \in [\omega_n(0),\omega_n(\pi)],
\end{equation}
\item If $\omega_n(0)=\omega_{max}$  and $\omega_n(\pi)=\omega_{min}$  of the band
\begin{equation}
\pderuno{\theta}{\omega}=-\left| \pderuno{\theta}{\omega}\right| <0,\qquad \forall \omega \in [\omega_n(0),\omega_n(\pi)].
\end{equation}
\end{itemize}
These two cases can be implemented in the same formula by taking the module of the Jacobian in (\ref{eq25}) as follows
\begin{equation}\label{eq27}
\bigtriangleup_T \mathcal{F}=  \sum_n \int_{\omega_n^{min}}^{\omega_n^{max}} \frac{d\omega}{ \pi}  \left|\pderuno{\theta}{\omega}\right| B(\omega,T).
\end{equation}
The band structure of the comb ensures that the allowed bands are those in which $\textrm{Re} \, [ \theta(\omega)]$ is non-zero. Hence, since ${\rm Re}[\theta(\omega)]$ is identically zero for the energies in forbidden bands, summing over the allowed bands is the same as integrating $\omega$ from 0 to $\infty$. In this way, the temperature  dependent part of the free energy takes the form
\begin{eqnarray}\label{eq26}
\bigtriangleup_T \mathcal{F}&=&  \int_{0}^{\infty} \frac{d\omega}{ \pi}\left| \textrm{Re} \left(\pderuno{\theta}{\omega} \right) \right|B(\omega,T)=\nonumber\\
&=& \int_{0}^{\infty} \frac{d\omega}{ \pi}\tilde{f}_\theta(\omega) B(\omega,T).
\end{eqnarray}
This fact allows us to give a general expression for the density of states of the comb as a function of {\color{black}$\omega$}
\begin{equation}\label{eq31}
\tilde{f}_q(\omega)= a \left|Re\left(\pderuno{q}{\omega}\right) \right|.
\end{equation}
The secular equation (\ref{eq6}) can be written in terms of the scattering transmission amplitude $t(k)$ and the phase shift $\delta(k)$ as
\begin{equation}\label{eq21}
\cos(\theta)= \frac{1}{|t(\omega)|} \cos(\omega a + \delta(\omega) + p\pi), \quad p\in \mathbb{Z},
\end{equation}
where the freedom introduced by the integers $p$ corresponds to the fact that the argument of the transmission scattering amplitude $t(k)$ is fixed by the arguments of the reflection amplitudes $r_{R,L}(k)$ up to $(2p+1)\pi$
\begin{equation*}
2{\rm arg}(t(k))={\rm arg}(r_R(k))+{\rm arg}(r_L(k))+(2p+1)\pi,\,\,\, p\in\mathbb{Z}
\end{equation*}
as pointed out in Ref. \cite{gad1909.08603}. {\color{black}The limit $a\to\infty$ in \eqref{eq21} does not exist in general. Nevertheless $a\to\infty$ keeping $\theta$ fixed can be well understood from a physical point of view following the equivalence between combs and selfadjoint extensions of $\hat H_0\equiv-d^2/dx^2$ in $[-a/2,a/2]$ defined by quasi-periodic boundary conditions shown in Ref. \cite{bord19-7-38}. The comb can be understood as the selfadjoint extension of the hamiltonian
\begin{equation*}
\hat H_V=-\frac{d^2}{d x^2}+V(x)
\end{equation*}
defined over the finite interval $[-a/2,a/2]$ with quasi-periodic boundary conditions. Since $V(x)$ has compact support in the interior of $[-a/2,a/2]$ any obstruction for $\hat H_V$ to be selfadjoint comes from $\hat H_0$, so the selfadjoint extensions of $\hat H_V$ and $\hat H_0$ are the same. Under these conditions when $a\to\infty$ the operator $\hat H_0$ (equivalently $\hat H_V$) becomes selfadjoint. Hence $a\to\infty$ gives rise to a quantum mechanical system defined by one single potential $V(x)$ with compact support over the whole real line. In this case from Ref. \cite{bord1807.10354} 
\begin{eqnarray}\label{eq33}
\bigtriangleup_T \mathcal{F}&=&  \int_{0}^{\infty} \frac{d\omega}{ \pi} \pderuno{\delta(\omega)}{\omega} B(\omega,T).
\end{eqnarray}
Therefore comparing Eq. \eqref{eq33} with Eqs. \eqref{eq26} and \eqref{eq31} we can give $f_\theta$ the physical meaning of a {\it phase shift} for particles propagating along the comb. In addition it is a well known fact that the derivative of the phase shift with respect $\omega$ is the density of states for the continuous spectrum defined by $\hat H_V$ as a quantum Hamiltonian over the real line which completes the physical analogy.
}

\subsection{Free energy: Matsubara formalism}
\newcommand{\suml}{\sum_{\ell=0}^{\infty}{\vphantom{\sum}}^{\prime}}
In the preceding subsection we have given a representation of the thermodynamic quantities for a finite temperature scalar quantum field theory in terms of real frequencies (or one particle energies). The Matsubara representation is an alternative to the one given above in terms of imaginary frequencies, $\omega \to i \xi$, that take discrete values $\xi_\ell=2\pi \ell T$, being $\ell$ an integer for bosons and half integer for fermions. The Matsubara representation can be obtained starting from an Euclidean field theory on a finite time interval \cite{bord-book}. Equivalently {\color{black}it arises} from the representation in terms of real frequencies  by {\color{black} performing} a Wick rotation. In this subsection we will obtain the Matsubara representation {\color{black}in the} case where the one particle spectrum has a band structure.  

{\color{black} In order to take $\alpha=\pi/2$ in (\ref{eq21a}) we must introduce a displacement $\epsilon>0$ on the resulting vertical line to avoid the branch points of $B(k,T)$, i.e., we turn the upper half of the contour towards the vertical semi-line $\omega=\epsilon+i\xi$ and the lower half of the contour towards $\omega=\epsilon-i\xi$. Before taking the limit $\epsilon\to 0$ the singular terms cancel, as explained in Ref. \cite{klaus-book}, and we finally obtain:}
\begin{eqnarray}\label{eq34}
&&\!\!\!\! \bigtriangleup_T \mathcal{F}=\int_{0}^{\pi} \frac{d\theta}{\pi}  \int_0^\infty \frac{d\xi}{2\pi i} \left[ - B(i\xi, T) \partial_\xi \log f_\theta(i \xi ) +\right.\nonumber\\
&&\left. \hspace{3.5cm} B(-i\xi, T) \partial_\xi \log f_\theta(-i \xi)\right],
\end{eqnarray}
where
\begin{eqnarray}
B(\pm i\xi, T)&=& T \log (1-e^{\mp i\xi /T}),\nonumber\\
f_\theta(\pm i\xi)&=& \cos \theta - h_V(\pm i\xi).
\end{eqnarray}
Further we use
{\color{black}\begin{eqnarray}\label{eq36}
\log (1-e^{\mp i\xi/T})& =& \log \left|2 \sin \frac{\xi}{2T} \right|\nonumber\\
&\mp& \frac{i\xi}{2T}   \pm i \pi \suml \Theta (\xi-\xi_\ell),\label{eq36}
\end{eqnarray}}
where $\Theta$ is the step function, the prime on the sum means that the  contribution from $\ell=0$ enters with a factor $1/2$ 
and $\xi_\ell= 2\pi T \ell$ ($ \ell \in \mathbb{Z}$) are the Matsubara frequencies. Since $f_\theta(i\xi)=f_\theta(-i\xi)$ for the potentials we are studying\footnote{{\color{black}We can write the scattering data using a common denominator which is basically the Jost function: $t(k)=\tau(k)/j(k),\,\,r_{R,L}(k)=\rho_{R,L}(k)/j(k)$. Using this notation, $t^2-r_Rr_L=j^*(k)/j(k)$, and it is a very well known property of the Jost function that $j(-k)=j^*(k)$. In addition $\tau(-k)=\tau(k),\,\,\rho_R(-k)=\rho_L(k)$. These properties ensure $f_\theta(i\xi)=f_\theta(-i\xi)$ for all those potentials with compact support and time reversal symmetry (see e. g. \cite{galin-b} for more details).}}, if we insert (\ref{eq36}) into (\ref{eq34})  the contribution with $\log|2 \sin(\xi/2T)|$ cancel and we arrive at
\begin{eqnarray}\label{eq37a}
&&\bigtriangleup_T \mathcal{F}=\int_{0}^{\pi} \frac{d\theta}{\pi}  \int_0^\infty \frac{d\xi}{2\pi }\,  \xi \,  \partial_\xi \log f_\theta(i \xi )\nonumber\\
&&-\int_{0}^{\pi} \frac{d\theta}{\pi}  \int_0^\infty d\xi\,\,  T \,  \partial_\xi \log f_\theta(i \xi) \suml \Theta (\xi-\xi_\ell).
\end{eqnarray}
Here the first term is, with minus sign, the vacuum energy
\begin{eqnarray}\label{eq38a}
E_0&=&-\int_{0}^{\pi} \frac{d\theta}{\pi}  \int_0^\infty \frac{d\xi}{2\pi }\,  \xi \,  \partial_\xi \log f_\theta(i \xi ).
\end{eqnarray}
Since $\mathcal{F}=E_0 +\bigtriangleup_T \mathcal{F} $ we get
\begin{equation}\label{eq39}
\mathcal{F}= -\int_{0}^{\pi} \frac{d\theta}{\pi}  \int_0^\infty d\xi\,\,  T \,  \partial_\xi \log f_\theta(i \xi) \suml \Theta (\xi-\xi_\ell).
\end{equation}
Finally, we   integrate  {by parts} and arrive at
\begin{equation}\label{eq40a}
\mathcal{F}= T \sum_{\ell=0}^\infty   \int_{0}^{\pi} \frac{d\theta}{\pi}    \log f_\theta(i \xi_\ell)
\end{equation}
for the free energy per unit cell, which is a more conventional form of the Matsubara representation.
This way we observe the expected feature that the vacuum energy is the zero temperature limit of the free energy. It must be mentioned that in eq. (\ref{eq37a}) an ultraviolet divergence was introduced. The temperature dependent part of the free energy is finite, however the vacuum energy and the free energy have a divergence. Therefore we should have introduced a regularization in separation the contributions in (\ref{eq37a}). The vacuum energy for a generalized Dirac comb was calculated in \cite{bord19-7-38} with eq. (44) as final formula. Thereby a renormalization was performed by subtracting the contributions from the vacuum energies of the $\delta$-$\delta^\prime$ potentials taken separately.

\subsection{Introducing a mass term}

When we study a potential that generates a spectrum with bound states it is convenient to introduce a mass term in order to avoid instabilities. We consider the action of a scalar massive field $\phi(x)$ in (1+1)-dimensions
\begin{equation}
S(\phi)= \frac{1}{2} \int dx^2 \left[\partial_\mu\phi(x) \partial^\mu \phi(x) - (m^2+U(x)) \phi(x)^2 \right]
\end{equation}
where $U(x)$ is the general periodic potential with compact support considered.
The modes of this scalar field obey the Schr\"odinger equation, after Fourier transform,
\begin{equation}
\left( -\pder{}{x}{2} + U(x)\right) \phi_\omega(x)= (\omega^2 - m^2) \phi_\omega(x),
\end{equation}
being $\omega=\sqrt{k^2+m^2}$ the frequencies of the quantum field modes. The temperature dependent part of the free energy is given by
\begin{eqnarray}
\bigtriangleup_T \mathcal{F}= \!\!\!\!\!\!\!\sum_{k \,  \in\,  \textrm{spec}(\hat{K})} \!\!\! \!\!B(\omega (k), T)\!=\!\!\!\!\!\!\!\!\sum_{k \,  \in\,  \textrm{spec}(\hat{K})}\!\!\!\! B(\sqrt{k^2+m^2}, T).
\end{eqnarray}
This last expression will consist of a summation over bound states ($k=i\kappa$ with $\kappa>0$) of the Boltzmann factor and a Cauchy integral over the states of the continuous spectrum (notice that the secular equation $f_\theta$ is holomorphic in $k$, not in $\omega$),
\begin{eqnarray}
&&\!\!\bigtriangleup_T \mathcal{F}\! = T\sum_{n} \log\left( 1- e^{\frac{\sqrt{m^2-\kappa_n^2}}{T}}\right) \nonumber\\
&& + \int_{0}^{\pi} \!\frac{d\theta}{\pi}\!\! \oint_\Gamma \frac{dk}{2\pi i} B(\sqrt{k^2+m^2},T) \partial_k \log f_\theta(k),
\end{eqnarray}
where $\Gamma$  is the contour represented in Figure \ref{fig:contour} but displaced a distance $m$ on the real axis. Again the integral on all the contour is reduced to the integral on the  lines $k=\xi e^{\pm i \alpha}$ with $\alpha$ a constant angle:
\begin{eqnarray}\label{eq44c}
&&\bigtriangleup_T \mathcal{F}= T\sum_{n} \log\left( 1- e^{\frac{\sqrt{m^2-\kappa_n^2}}{T}}\right) \nonumber\\
&&+ \int_{0}^{\pi} \frac{d\theta}{\pi} \int_m^\infty \frac{d\xi}{2\pi i} \left( B(z(\xi,-\alpha,m),T)\partial_k \log f_\theta(\xi e^{-i\alpha})\right. \nonumber\\
&&\left. - B(z(\xi,\alpha,m),T)  \partial_k \log f_\theta(\xi e^{i\alpha})  \right).
\end{eqnarray}
where we have written $z(\xi,\alpha,m)=\sqrt{\xi^2 e^{i2\alpha}+m^2}$.
It is of note that the bound states are poles in the pure imaginary axis of the $k$-complex plane. Hence, the integral in (\ref{eq44c}) is well defined if we use a contour similar to that represented in Figure \ref{fig:contour}.

\section{Particular cases}
%
In this section we apply the formulas for the free energy and the entropy developed in the preceding section to some specific systems. The first is a single potential with point support, the other two have periodic potentials, one with localized support and the other with extended support (within one lattice cell).

\subsection{Entropy for single $\boldsymbol{\delta\text{-}\delta^\prime}$ potential}
We consider the potential $$V_{\delta \delta^{\prime}}=w_0 \delta(x)+2w_1 \delta^{\prime}(x),$$ {\color{black}with $w_0>0$ to ensure that there are no negative energy levels, and in addition no negative energy bands in the associated comb (see Ref. \cite{gadella-pla09} for a detailed discussion about the bound state spectrum of the $\delta$-$\delta'$ potential).} For this potential, following \cite{muno15-91-025028}, the scattering amplitudes are given by
\begin{eqnarray}
&&\qquad \qquad \qquad t(k)=-\frac{2k(w_1^2-1)}{2k(w_1^2+1)+iw_0},\nonumber\\
&&r_R(k)=\frac{-4 k w_1-i w_0}{2 k \left(w_1^2+1\right)+i w_0},\,\, r_L(k)=\frac{4 k w_1-i w_0}{2 k \left(w_1^2+1\right)+i w_0}.\nonumber
\end{eqnarray}
Taking into account that the determinant of the scattering matrix is $t^2-r_R r_L=e^{2 i \delta(k)}$ {\color{black} being $\delta(k)$ the phase shift, we obtain}
\begin{equation}
\tan(2\delta(k))=\frac{4 k/\gamma}{1-4 k^2/\gamma^2},\quad \text{with}\quad \gamma\equiv\frac{w_0}{1+w_1^2}
\end{equation}
Transforming the $\arctan$ in the last equality {\color{black} by means of} the trigonometric relation {\color{black}$$\tan (2z)= \frac{2 \tan z}{ 1-\tan^2 z}$$} {\color{black} we} obtain two possible solutions for $\tan(\delta(k))$:
\begin{equation}
\tan(\delta(k))=\begin{cases}
 +2k/\gamma, \\
 -\gamma/2k.
\end{cases}
\end{equation}
If we require that $\delta(k\to\infty)=0$ then the only possibility is $\tan(\delta(k))=-\gamma/2k$. Taking now the $\arctan$ and using
\begin{equation}
\arctan(-z)=-\arctan(z),
\end{equation}
\begin{equation}
\arctan\left(\frac{1}{z}\right)=\begin{cases}
 \frac{\pi}{2}-\arctan(z), & (z>0), \\
 -\frac{\pi}{2}-\arctan(z), & (z<0),
\end{cases}
\end{equation}
we finally obtain the expression for the phase shift,
\begin{equation}\label{eq38}
\delta(k)=\begin{cases}
 -\frac{\pi}{2}+\arctan\left(\frac{2k}{\gamma}\right), & (\gamma>0), \\
  +\frac{\pi}{2}+\arctan\left(\frac{2k}{\gamma}\right), & (\gamma<0).
\end{cases}
\end{equation}
Using the derivative of the phase shift (\ref{eq38}),
\begin{equation}
 \pderuno{}{k} \delta(k)= \frac{2w_0(1+w_1^2)}{w_0^2+4k^2(1+w_1^2)^2},
\end{equation}
in  (\ref{eq33}), the free energy can be evaluated when $w_0> 0$ (and therefore there are no bound states) or the binding energy should be larger than the mass. In this way, the temperature dependent part of the free energy and the entropy can be written explicitly as
\begin{equation}
\bigtriangleup_T \mathcal{F} =\frac{T}{\pi} \int_0^\infty dk  \frac{2w_0(1+w_1^2)\ln \left[ 1-e^{-{k}/T} \right]}{w_0^2+4k^2(1+w_1^2)^2},\label{eq40}
\end{equation}
\vspace{0.1cm}
\begin{eqnarray}
S&=&\frac{1}{\pi} \int_0^\infty dk \left[ -\ln  \left(1-e^{-{k}/T} \right)+\frac{ k/T}{e^{{k}/T}-1}   \right] \times\nonumber\\
&&\hspace{2cm}\times\frac{2w_0(1+w_1^2)}{w_0^2+4k^2(1+w_1^2)^2}.\label{eq41}
\end{eqnarray}
\begin{figure}[h]
\centering
\includegraphics[width=0.485\textwidth]{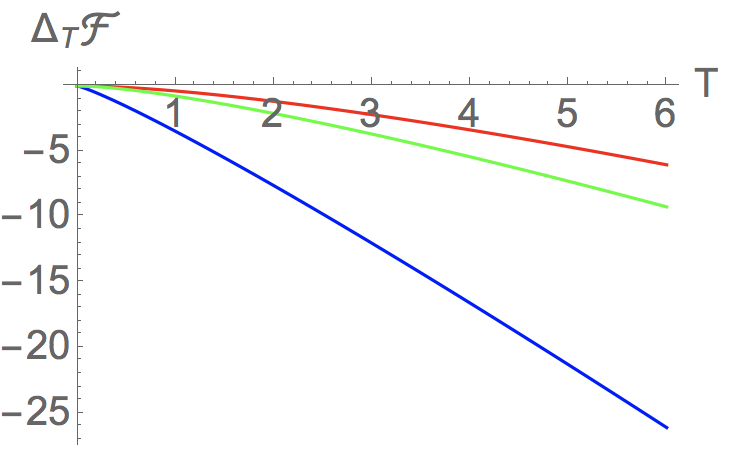} 
\caption{\small Temperature dependent part of the free energy (\ref{eq40}), for a single $\delta\text{-}\delta'$-potential as a function of $T$ for the potential $V_{\delta \delta^{\prime}}$ and the configurations $w_0=0.01$,$w_1=2$ (blue); $w_0=3$,$w_1=2$  (green) and $w_0=2$,$w_1=0$  (red).}
\label{fig:freesingle}
\end{figure}
These formulas generalize Eq. (4) from Ref. \cite{bord1807.10354} to the case $w_1\ne0$.
The above two expressions can easily be evaluated numerically. Results are shown in Figures \ref{fig:freesingle} and \ref{fig:freesingleS}. As it can be seen, in all cases the entropy is positive.
\begin{figure}[h]
\centering
\includegraphics[width=0.485\textwidth]{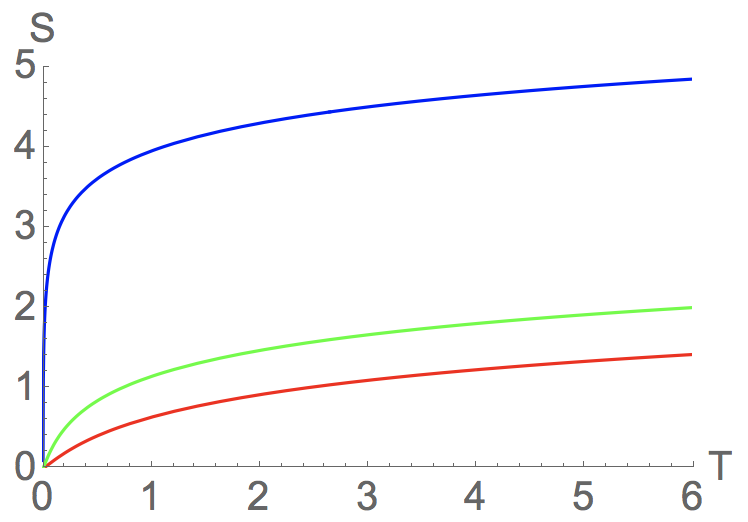}
\caption{\small Entropy  (\ref{eq41}), for a single $\delta\text{-}\delta'$-potential as a function of $T$ for the potential $V_{\delta \delta^{\prime}}$ and the configurations $w_0=0.01$,$w_1=2$ (blue); $w_0=3$,$w_1=2$  (green) and $w_0=2$,$w_1=0$  (red).}
\label{fig:freesingleS}
\end{figure}


\subsection{Entropy for the Dirac comb}
\begin{figure}[h!]
\centering
\includegraphics[width=0.485\textwidth]{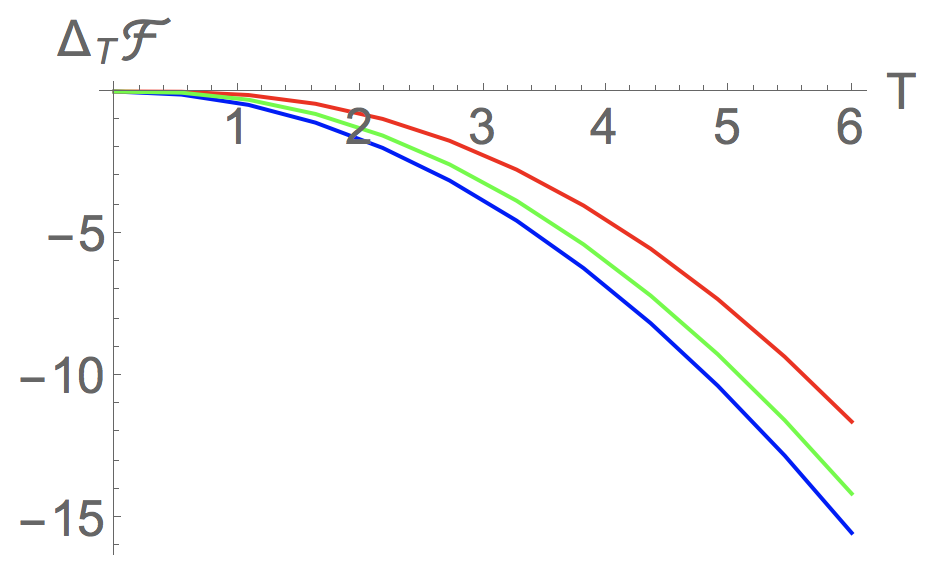}
\vspace{-0.2cm}\caption{\small Free energy $\bigtriangleup_T  \mathcal{F}$, (\ref{eq21a}), as a function of $T$ for the potential (\ref{eq52}) for $w_0=0.1, w_1=5$ (blue), $w_0=8, w_1=0$ (red), $w_0=3, w_1=2$ (green). We have chosen for the lattice spacing $a=1$.}
\label{fig:freecomb1L}
\end{figure}

As potential we take a periodic chain of $\delta$-$\delta^\prime$ functions \Ref{1},
\begin{equation}\label{eq52}
V(x)= \sum_{n=-\infty}^\infty\left( w_0 \delta(x-an)+ 2w_1 \delta^\prime(x-an)\right),
\end{equation}
with lattice spacing $a$. This model is a generalization of the \textit{Dirac comb model}. Changing notations for convenience \cite{bord19-7-38}, the spectral equation \Ref{eq22} reads
\eq{eq3}{
&& g_q(\omega)\equiv\Omega \cos(q a)+ \cos(\omega a)+\frac{\gamma}{2\omega} \sin(\omega a) =0
}
with
\eq{eq3a}{
 \hspace{1.5cm} \gamma&=\frac{w_0}{w_1^2+1}, \qquad \Omega= \frac{w_1^2-1}{w_1^2+1}.
}
For the $\delta$-$\delta'$ comb all we need to do is use the momentum representation formula for the temperature dependet part of the free energy (\ref{eq21a}).
From  the expression (\ref{eq3}) it is easy to see that in this case
\begin{equation}\label{eq51}
h_V(k)= \frac{-1}{\Omega} \left( \cos(ka) + \frac{\gamma}{2k}\sin(ka) \right).
\end{equation}
On the one hand, plugging (\ref{eq51})  in (\ref{eq21a}) we can calculate the thermal correction to the free energy of the comb  at any temperature. Figure \ref{fig:freecomb1L} shows the temperature dependent part of the free energy for different configurations of a $\delta$-$\delta^\prime$ comb as a function of temperature.

\begin{figure}[h!]
\centering
\includegraphics[width=0.48\textwidth]{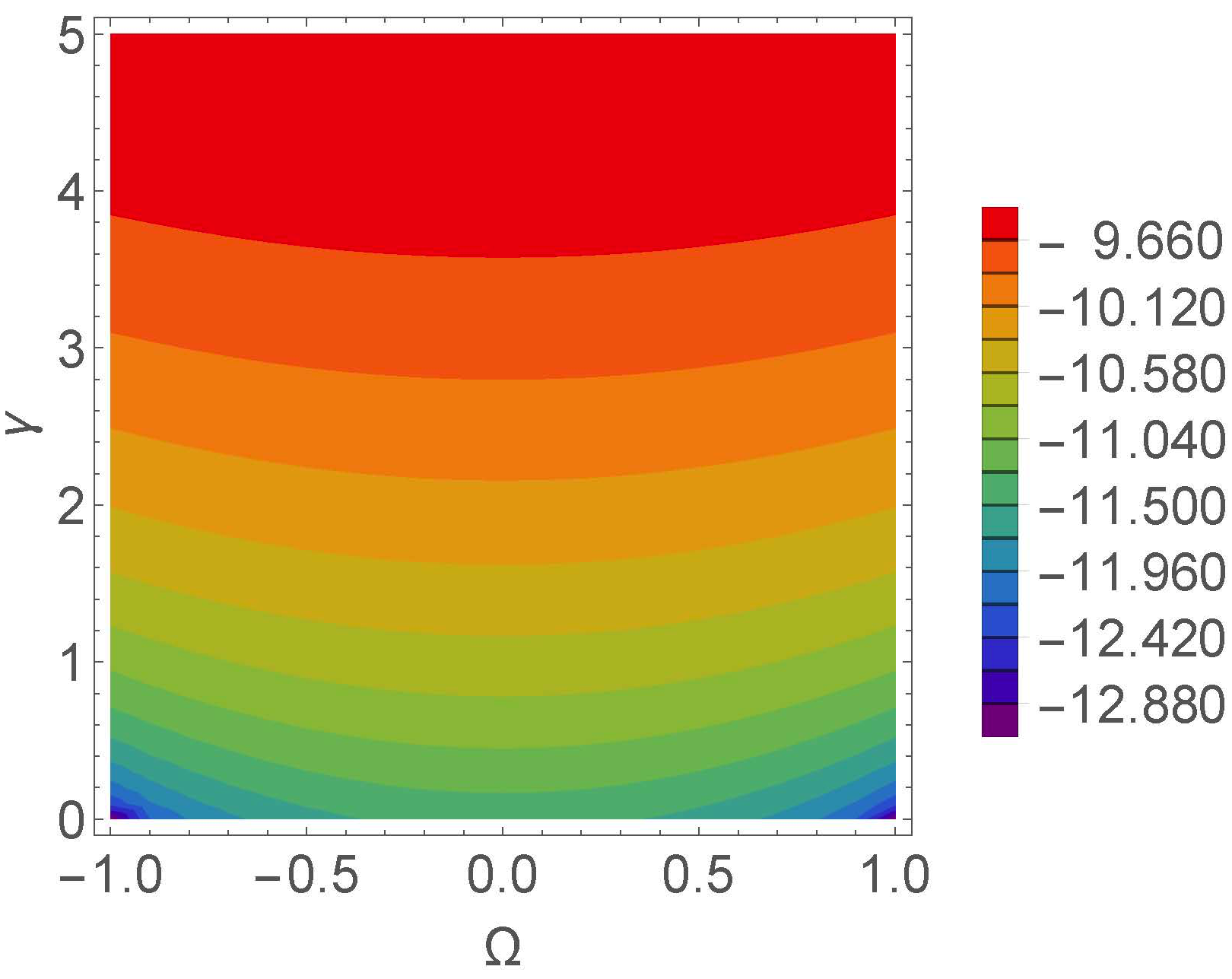}
\caption{\small Free energy $\bigtriangleup_T  \mathcal{F}$, (\ref{eq21a}) for $T=5$ for the potential (\ref{eq52}) in the parameter space $\Omega-\gamma$. We have chosen for the lattice spacing $a=1$.}
\label{fig:freecomb2TH}
\end{figure}

\begin{figure}[h!]
\centering
\includegraphics[width=0.48\textwidth]{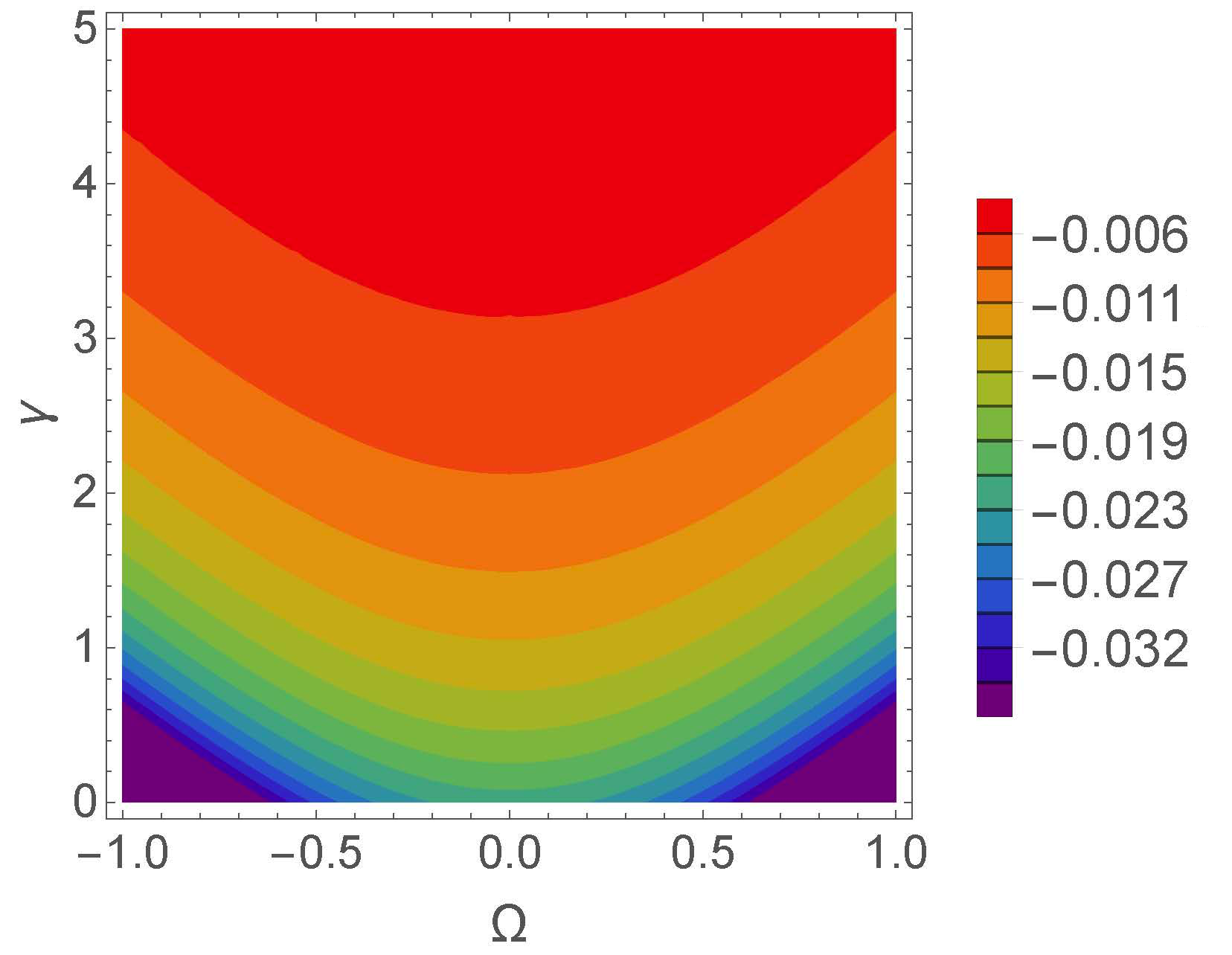}
\caption{\small Free energy $\bigtriangleup_T  \mathcal{F}$, (\ref{eq21a}) for $T=0.5$  for the potential (\ref{eq52}) in the parameter space $\Omega-\gamma$. We have chosen for the lattice spacing $a=1$.}
\label{fig:freecomb2TL}
\end{figure}

 \begin{figure}[h!]
\centering
\includegraphics[width=0.485\textwidth]{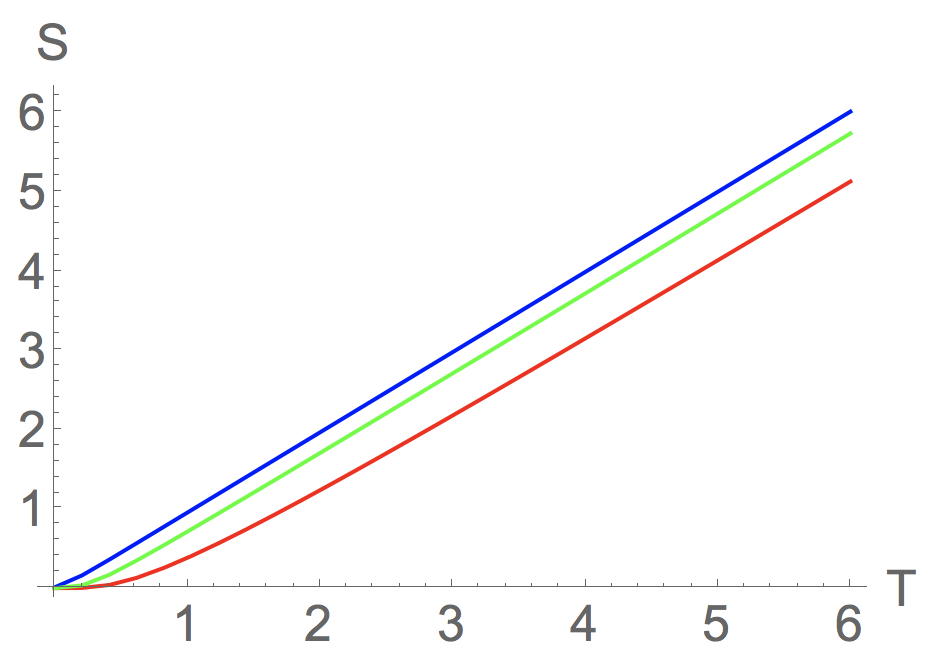}
\vspace{-0.2cm}\caption{\small Entropy as functions of $T$ for the potential (\ref{eq52}) for $w_0=0.1, w_1=5$ (blue), $w_0=8, w_1=0$ (red), $w_0=3, w_1=2$ (green). We have chosen for the lattice spacing $a=1$.}
\label{fig:freecomb1R}
\end{figure}
Plots in Figures \ref{fig:freecomb2TL} and \ref{fig:freecomb2TH} show the thermal correction to the free energy in the parameter space $\Omega-\gamma$ in the regimes of high and low temperatures respectively. In both cases, $\bigtriangleup_T  \mathcal{F}$ takes negative values. In the limit of low temperatures, we can see that the leading contribution to the free energy will be provided for the vacuum energy at zero temperature, whereas the thermal correction will be a small deviation as it should be. However, in the limit of high temperatures  the opposite happens and the thermal correction becomes more important.
\begin{figure}[h]
\centering
\includegraphics[width=0.485\textwidth]{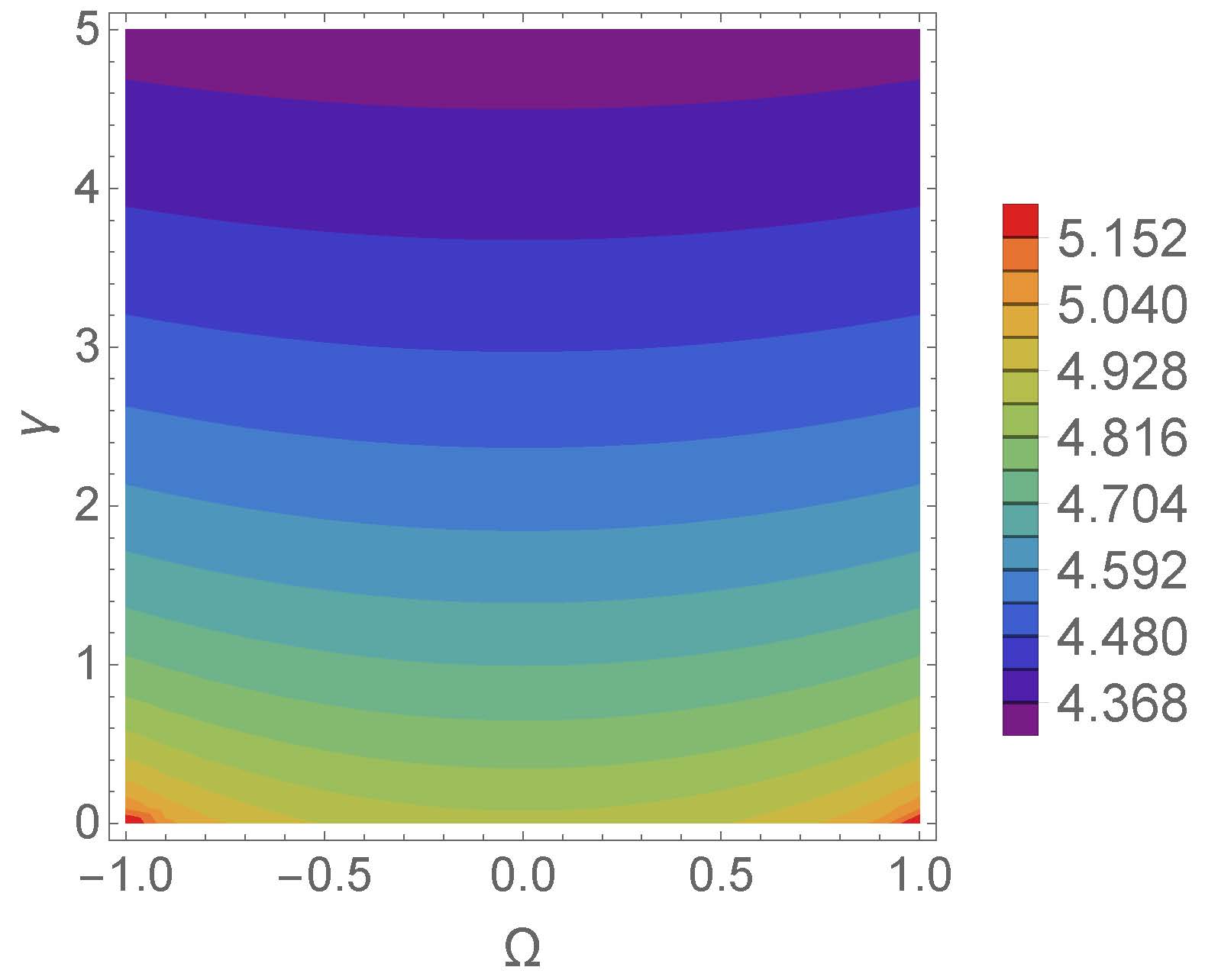}
\caption{\small The entropy $S$ for $T=5$ for the $\delta$-$\delta^\prime$ comb, (\ref{eq52}), in the parameter space $\Omega-\gamma$. We have chosen for the lattice spacing $a=1$.}
\label{fig:entcomb2HT}
\end{figure}
\begin{figure}[h]
\centering
\includegraphics[width=0.485\textwidth]{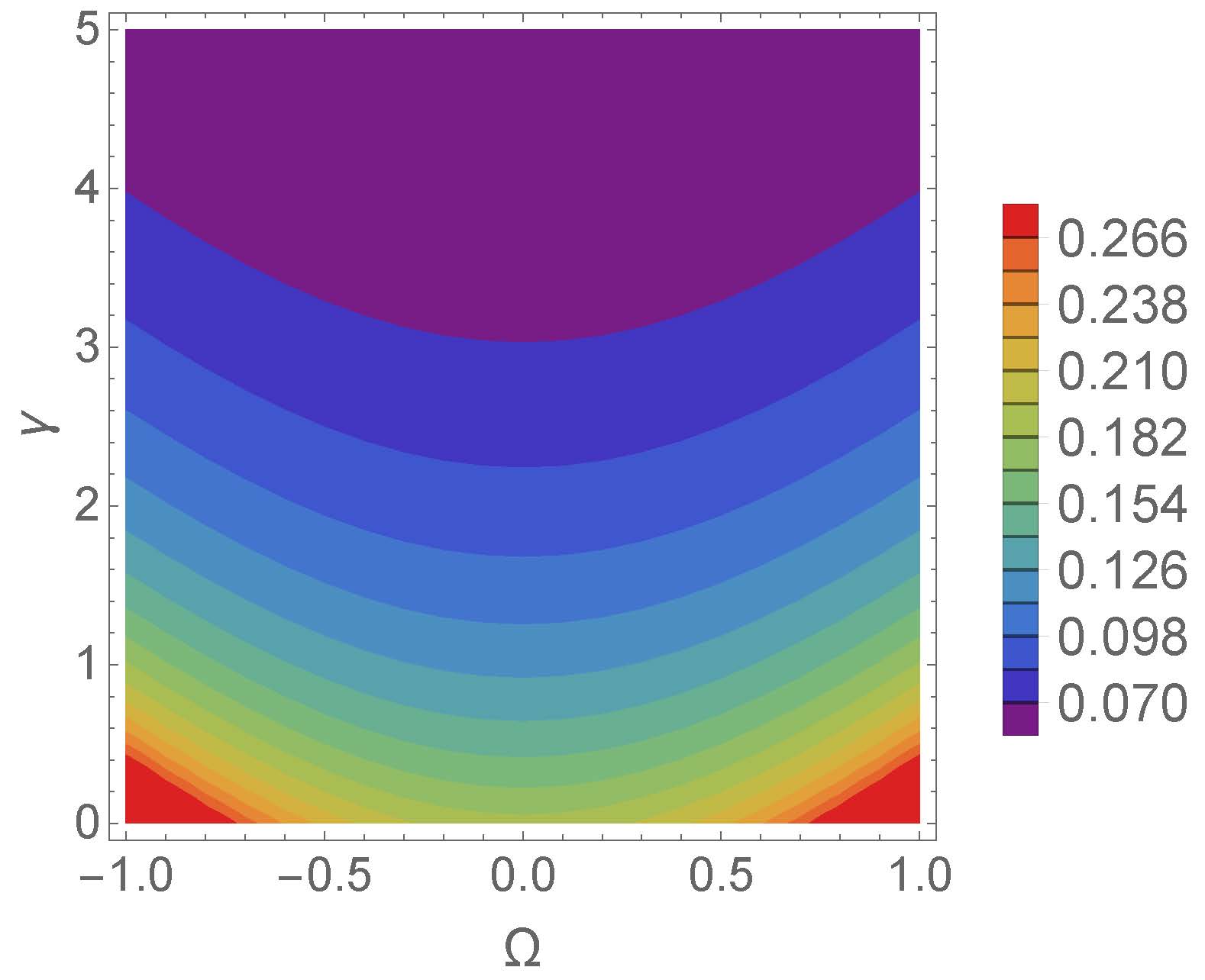}
\caption{\small The entropy $S$ for $T=0.5$ for the $\delta$-$\delta^\prime$ comb, (\ref{eq52}), in the parameter space $\Omega-\gamma$. We have chosen for the lattice spacing $a=1$.}
\label{fig:entcomb2LT}
\end{figure}

On the other hand, if we derive eq. (\ref{eq21a}) with respect to the temperature and we change the global sign, we obtain the entropy of the comb, which can be evaluated at any finite non-zero temperature for different configurations of a $\delta$-$\delta^\prime$ comb (Figure \ref{fig:freecomb1R}).
Plots in Figures \ref{fig:entcomb2HT} \ref{fig:entcomb2LT}  show the behaviour of the entropy in the parameter space $\Omega-\gamma$ in the regimes of high and low temperatures respectively. In both cases the entropy $S$ takes positive values {\color{black}for any value of the temperature} as can be seen in Figure \ref{fig:freecomb1R}.

\subsection{Entropy for sine-Gordon comb}
\begin{figure}[h]
\centering
\includegraphics[width=0.48\textwidth]{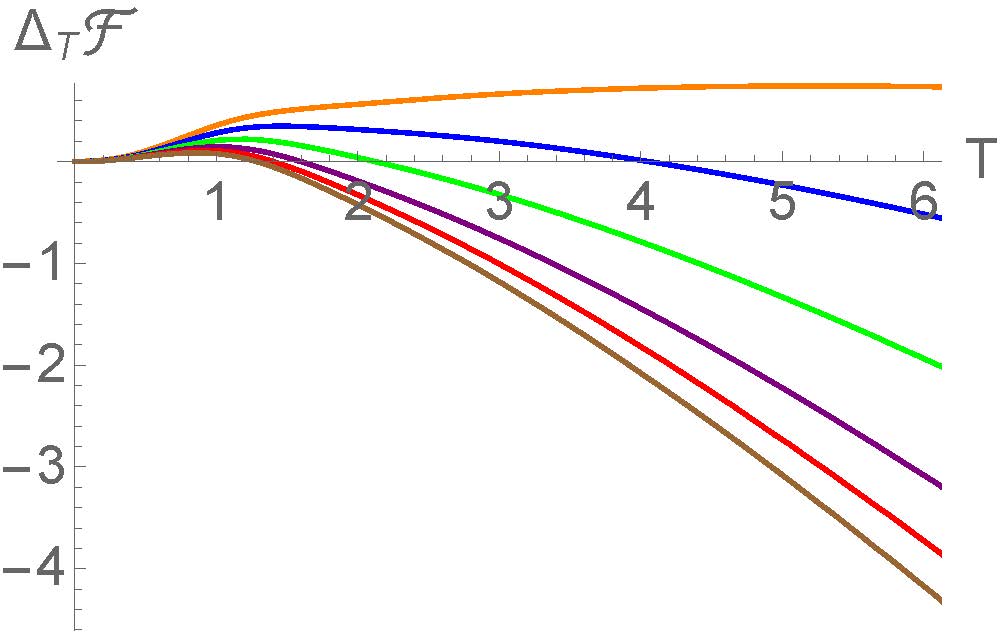}
\vspace{-0.2cm}\caption{Free energy, $\bigtriangleup_T  \mathcal{F}$, (\ref{eq21a}), for the 'sine-Gordon' comb as a function of $T$ for: $\epsilon=0.1$ (orange line), $\epsilon=0.25$ (blue line), $\epsilon=0.5$ (green line), $\epsilon=0.75$ (purple line), $\epsilon=0.9$ (red line) and $\epsilon=1$ (brown line).  We have chosen for the lattice spacing $a=1$.}
\label{fig:freekink3}
\end{figure}
As potential we take a periodic chain of sine-Gordon kinks, represented by P\"oschl-Teller potentials as follows,
\begin{eqnarray}\label{eq52b}
U_{PT}(x)&=& \sum_{n=-\infty}^\infty V_{PT,\epsilon}(x-na),\\
V_{PT,\epsilon}(x)&=&1- \frac{2\Theta(-x+\epsilon/2)\Theta(x+\epsilon/2)}{\cosh^2 (x)},
\end{eqnarray}
being $\Theta$ the Heaviside step function, $\epsilon$ the length of the compact support of the potential and $a$ the lattice spacing (notice that $0<\epsilon\leq a$). Following \cite{guil11-50-2227}, the scattering coefficients are
\begin{eqnarray}
t(k)&=&\frac{4k^2(k^2+1)}{\Delta(k)},\\
r(k)&=& \frac{e^{i\epsilon k}\Lambda(\Lambda + 2k (k+i\tanh (\epsilon /2)))}{\Delta(k)}-\nonumber\\
&&\frac{e^{-i\epsilon k}\Lambda(\Lambda + 2k (k-i\tanh (\epsilon /2)))}{\Delta(k)},\nonumber\\
\Delta (k) &=&-e^{2i\epsilon k} \Lambda^2 + [\Lambda+2k(k-i\tanh (\epsilon /2))]^2\nonumber
\end{eqnarray}
with $\Lambda=1-\tanh^2 (\epsilon/2)$. The poles of the determinant of the scattering matrix  ($\textrm{det} S= t^2-r_Rr_L$) are the bound states ($k=i\kappa$ with $\kappa>0$) of the kink-comb spectrum. In this case we find that there are no bound states {\color{black}(see Ref. \cite{guil11-50-2227})}.

In this lattice, the bands are determined by the real solutions of the spectral equation, which takes the form
\eq{eq3d}{
\tilde{g}_q(k)&= \cos(q a) - h_V(k)
}
with the functions
\eq{eq3c}{
&h_V(k)= \frac{\Sigma \cos(ka)-\Upsilon \sin(ka)}{k^2(k^2+1)(1+\cosh \epsilon)} ,\\
&\Upsilon= 2k \tanh (\frac{\epsilon}{2})(1+k^2+k^2 \cosh(\epsilon)) +  \Lambda^2 \cos(k\epsilon) \sin(\epsilon),\nonumber\\
 &\Sigma= k^2(3+k^2)+k^2(-1+k^2)\cosh (\epsilon) + \Lambda^2 \sin^2(k\epsilon).\nonumber
}

\begin{figure}[h]
\centering
\includegraphics[width=0.48\textwidth]{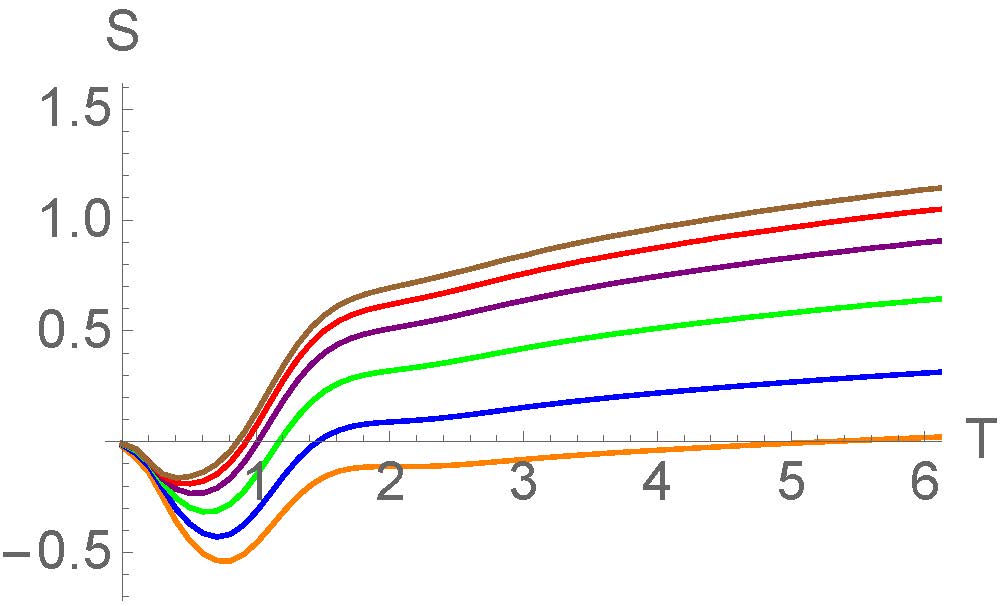}
\vspace{-0.2cm}\caption{Entropy for the 'sine-Gordon' comb as a function of $T$ for: $\epsilon=0.1$ (orange line), $\epsilon=0.25$ (blue line), $\epsilon=0.5$ (green line), $\epsilon=0.75$ (purple line), $\epsilon=0.9$ (red line) and $\epsilon=1$ (brown line).  We have chosen for the lattice spacing $a=1$.}
\label{fig:freekink3S}
\end{figure}

For the kink-comb all we need to do is use the momentum representation formula for the free energy (\ref{eq21a}) being (\ref{eq3d}) the spectral equation in this case. The result is shown in Figure  \ref{fig:freekink3} as a function of temperature for different values of the compact support length.  By deriving the free energy with respect to the temperature, the entropy of the system is obtained and can be evaluated for any non-zero finite temperature. The result is shown in Figure \ref{fig:freekink3S}. It can be seen that there are negative entropies if the kink's compact support is such that $\epsilon < a$. Even in the limit of a continuous comb, i. e. {\color{black}$\epsilon=a$} (brown line in the right plot of Fig. \ref{fig:freekink3S}), there are negative entropies.

\section{Conclusions}
In the foregoing sections we considered   free energy and entropy for periodic lattices built from infinite arrays of potentials with compact support. We have considered the particular cases of $\delta$-$\delta'$ potential, \Ref{eq52}, in one case and a P\"oschl-Teller potential, \Ref{eq52b}, in the other. First we derived some general representations for the \td quantities for a scalar quantum field theory with a classical background given by a generic periodic potential. The most commonly used is in terms of real frequencies and the Boltzmann factor. It is, so to say, the most physical one and convenient  due to the exponential decrease for large frequencies in its temperature dependent part. However, in dependence of the mode density, it may involve large oscillations. By turning the integration contour towards the imaginary axis by a finite angle $\al$ ($\al<\pi/2$, see Fig. 1), a still exponentially  convergent representation \Ref{eq26} appears having the advantage that it avoids large oscillations. Turning the contour finally to the imaginary axis, $\al=\pi/2$, we come to the Matsubara representation, \Ref{eq40a}, which this way is applicable also to a spectral problem with band structure.

For numerical evaluation, the intermediate representation \Ref{eq26} is most convenient. We used it for calculating free energy and entropy for the mentioned systems. For the generalized $\delta$-$\delta'$ comb we obtain a positive entropy, generalizing earlier results. Nevertheless for the periodic array of truncated P\"oschl-Teller potentials we obtained for temperatures below a certain value a negative entropy, generalizing earlier results in \cite{bord1807.10354} for a single plasma point on a half axis. These negative entropy regimes survive even in the continuum limit (brown line in the right plot of Figure \ref{fig:freekink3}. As discussed in Ref. \cite{thir70-235-339} the appearance of negative entropies can be a hint of instabilities of the quantum system.

It must be mentioned that so far no general rule can be guessed for the sign of the entropy calculated the way as in this and earlier papers. More work in this direction seems necessary in order to understand which are the fundamental properties that determine the sign of the entropy in quantum field theories under the influence of classical backgrounds.

\begin{acknowledgements}
The authors are grateful to the Spanish Government-MINECO (MTM2014- 57129-C2-1-P) for the financial support received. JMMC and LSS are grateful to the {\it Junta de Castilla y Le\'on} (BU229P18, VA137G18 and VA057U16) for the financial support. LSS is grateful to the Spanish Government-MINECO for the FPU-fellowships programme (FPU18/00957).
\end{acknowledgements}



\begin{thebibliography}{10}

\bibitem{cas48}
H.~B.~G. Casimir.
\newblock {On the Attraction Between Two Perfectly Conducting Plates}.
\newblock {\em Indag. Math.}, 10:261--263, 1948.
\newblock [Kon. Ned. Akad. Wetensch. Proc.100N3-4,61(1997)].

\bibitem{spa57}
M.~J. Sparnaay.
\newblock {Attractive Forces between Flat Plates}.
\newblock {\em Nature}, 180:334–335, 1957.

\bibitem{spa58}
M.~J. Sparnaay.
\newblock {Measurements of attractive forces between flat plates}.
\newblock {\em Physica}, 24:751--764, 1958.

\bibitem{Grib1994}
A.A. Grib, S.G. Mamayev, and V.M. Mostepanenko.
\newblock {\em Vacuum Quantum Effects in Strong Fields}.
\newblock Friedmann Laboratory Publishing, St. Petersburg, 1994.

\bibitem{milt-book}
K~A Milton.
\newblock {\em The Casimir effect: Physical manifestations of zero-point
  energy}.
\newblock WORLD SCIENTIFIC, 2001.

\bibitem{bord-book}
M.~Bordag, G.~L. Klimchitskaya, U.~Mohideen, and V.~M. Mostepanenko.
\newblock {\em {Advances in the Casimir effect}}.
\newblock Int. Ser. Monogr. Phys. Oxford University Press, 2009.

\bibitem{emig-prl07}
T.~Emig, N.~Graham, R.~L. Jaffe, and M.~Kardar.
\newblock {Casimir forces between arbitrary compact objects}.
\newblock {\em Phys. Rev. Lett.}, 99:170403, 2007.

\bibitem{emig-prd08}
T.~Emig, N.~Graham, R.~L. Jaffe, and M.~Kardar.
\newblock {Casimir Forces between Compact Objects. I. The Scalar Case}.
\newblock {\em Phys. Rev.}, D77:025005, 2008.

\bibitem{rahi-ped09}
Sahand~Jamal Rahi, Thorsten Emig, Noah Graham, Robert~L. Jaffe, and Mehran
  Kardar.
\newblock {Scattering Theory Approach to Electrodynamic Casimir Forces}.
\newblock {\em Phys. Rev.}, D80:085021, 2009.

\bibitem{kenneth-prb08}
Oded Kenneth and Israel Klich.
\newblock {Casimir forces in a T operator approach}.
\newblock {\em Phys. Rev.}, B78:014103, 2008.

\bibitem{kenneth-prl06}
Oded Kenneth and Israel Klich.
\newblock {Opposites attract: A Theorem about the Casimir force}.
\newblock {\em Phys. Rev. Lett.}, 97:160401, 2006.

\bibitem{asorey-npb13}
M.~Asorey and J.M. Mu{\~{n}}oz-Casta{\~{n}}eda.
\newblock Attractive and repulsive Casimir vacuum energy with general boundary
  conditions.
\newblock {\em Nucl. Phys. B}, 874(3):852 -- 876, 2013.

\bibitem{asorey-jpa06}
M~Asorey, D~Garc{\'{\i}}a {\'{A}}lvarez, and J~M Mu{\~{n}}oz-Casta{\~{n}}eda.
\newblock Casimir effect and global theory of boundary conditions.
\newblock {\em Journal of Physics A: Mathematical and General},
  39(21):6127--6136, 2006.

\bibitem{muno15-91-025028}
J.~M. Mu\~noz Casta\~neda and J.~Mateos~Guilarte.
\newblock ''$\delta\text{-}\delta'$ generalized Robin boundary conditions and
  quantum vacuum fluctuations''.
\newblock {\em Phys. Rev. D}, 91:025028, 2015.

\bibitem{munoz-lmp15}
Jose~M. Mu{\~{n}}oz-Casta{\~{n}}eda, Klaus Kirsten, and Michael Bordag.
\newblock Qft over the finite line. heat kernel coefficients, spectral zeta
  functions and selfadjoint extensions.
\newblock {\em Lett. Math.Phys.}, 105(4):523--549, 2015.

\bibitem{geye05-72-022111}
B.~Geyer, G.~L. Klimchitskaya, and V.~M. Mostepanenko.
\newblock Thermal corrections in the Casimir interaction between a metal and
  dielectric.
\newblock {\em Phys. Rev. A}, 72:022111, Aug 2005.

\bibitem{thir70-235-339}
W.~Thirring.
\newblock Systems with negative specific heat.
\newblock {\em Zeitschrift f{\"u}r Physik A Hadrons and nuclei},
  235(4):339--352, Aug 1970.

\bibitem{lium19-100-081406}
Mingyue Liu, Jun Xu, G.~L. Klimchitskaya, V.~M. Mostepanenko, and U.~Mohideen.
\newblock Examining the Casimir puzzle with an upgraded afm-based technique and
  advanced surface cleaning.
\newblock {\em Phys. Rev. B}, 100:081406, Aug 2019.

\bibitem{milt17-96-085007}
K.~A. Milton, Pushpa Kalauni, Prachi Parashar, and Yang Li.
\newblock {Casimir self-entropy of a spherical electromagnetic $\delta$
  -function shell}.
\newblock {\em Phys. Rev.}, D96(8):085007, 2017.

\bibitem{li16-94-085010}
Yang Li, K.~A. Milton, Pushpa Kalauni, and Prachi Parashar.
\newblock {Casimir Self-Entropy of an Electromagnetic Thin Sheet}.
\newblock {\em Phys. Rev.}, D94(8):085010, 2016.

\bibitem{bord18-51-455001}
M.~Bordag and K.~Kirsten.
\newblock {On the entropy of a spherical plasma shell}.
\newblock {\em J. Phys.}, A51(45):455001, 2018.

\bibitem{bord18-98-085010}
M.~Bordag.
\newblock {Free energy and entropy for thin sheets}.
\newblock {\em Phys. Rev.}, D98(8):085010, 2018.

\bibitem{bord1807.10354}
M.~Bordag.
\newblock Entropy in some simple one-dimensional configurations.
\newblock {\em arXiv:1807.10354}, 2018.

\bibitem{gadella-pla09}
M.~Gadella, J.~Negro, and L.M. Nieto.
\newblock Bound states and scattering coefficients of the
  $-a\delta(x)+b\delta'(x)$ potential.
\newblock {\em Phys.Lett. A}, 373(15):1310 -- 1313, 2009.


\bibitem{bord19-7-38}
M.~Bordag, J.M. Mu{\~{n}}oz-Casta{\~{n}}eda, and L.~Santamaria-Sanz.
\newblock Vacuum energy for generalised Dirac combs at $\textit{T}=0$.
\newblock {\em Front. Phys.}, 7, 2018.

\bibitem{guil11-50-2227}
J.~Mateos Guilarte and J.~M. Mu{\~{n}}oz-Casta{\~{n}}eda.
\newblock Double-delta potentials: one dimensional scattering. The Casimir effect and kink fluctuations.
\newblock {\em Int. J. Theor. Phys.}, 50(7):2227--2241,
  2011.
  
\bibitem{galin-b}
{\color{black}A.~Galindo and P.~Pascual
\newblock {\em {Quantum Mechanics I}}.
\newblock Springer-Verlag Berlin Heidelberg, 1990.}


\bibitem{gad1909.08603}
M.~Gadella, J.~M.~Mateos Guilarte, J.~M. Mu{\~{n}}oz-Casta{\~{n}}eda, L.~M.
  Nieto, and L.~Santamar\'\i a~Sanz.
\newblock Band spectra of periodic hybrid $\delta\text{-}\delta'$ structures.
\newblock {\em arXiv:1909.08603}, 2019.

\bibitem{klaus-book}
K. Kirsten.
\newblock {\em {Spectral Functions in Mathematics and Physics}}.
\newblock Chapman and Hall/CRC, 2001.


\end{thebibliography}


\end{document}